\documentclass[showpacs,preprintnumbers,amsmath,amssymb,floatfix]{revtex4}

\usepackage{color}   
\usepackage{graphicx}
\usepackage{dcolumn}
\usepackage{bm}

\begin{document}

\def\bbox#1{\hbox{\boldmath${#1}$}} 
\def\blambda{{\hbox{\boldmath $\lambda$}}} 
\def\eeta{{\hbox{\boldmath $\eta$}}}
\def\bxi{{\hbox{\boldmath $\xi$}}} 
\def\bzeta{{\hbox{\boldmath $\zeta$}}}
\def\sD{D \!\!\!\!/}

\title{ Anomalous Soft Photons in Hadron Production}

\author{Cheuk-Yin Wong}

\affiliation{Physics Division, Oak Ridge National Laboratory, 
Oak Ridge, TN\footnote{wongc@ornl.gov} 37831}

\affiliation{ Department of Physics and Astronomy, University of
Tennessee, Knoxville, TN 37996}

\date{\today}

\begin{abstract}

  Anomalous soft photons in excess of what is expected from
  electromagnetic bremsstrahlung have been observed in association
  with the production of hadrons, mostly mesons, in high-energy $K^+
  p$, $\pi^+ p$, $\pi^- p$, $pp$, and $e^+e^-$ collisions.  We propose
  a model for the simultaneous production of anomalous soft photons
  and mesons in quantum field theory, in which the meson production
  arises from the oscillation of color charge densities of the quarks
  of the underlying vacuum in the flux tube. As a quark carries both a
  color charge and an electric charge, the oscillation of the color
  charge densities will be accompanied by the oscillation of electric
  charge densities, which will in turn lead to the simultaneous
  production of soft photons during the meson production process.  How
  the production of these soft photons may explain the anomalous soft
  photon data will be discussed.  Further experimental measurements to
  test the model will be proposed.

\end{abstract}


\maketitle


\section{Introduction}

Anomalous soft photons are soft photons in excess of what is expected
from electromagnetic bremsstrahlung.  They have been observed in
conjunction with the production of hadrons, mostly mesons, in $K^+ p$
\cite{Chl84,Bot91}, $\pi^+ p$ \cite{Bot91}, $\pi^- p$
\cite{Ban93,Bel97,Bel02a}, $pp$ collisions \cite{Bel02}, and in
high-energy $e^+$-$e^-$ annihilations in $Z^0$ hadronic decay
\cite{DEL06,DEL09,Per09}.  Recent DELPHI measurements on the
characteristics of the produced hadrons associated with the anomalous
soft photon production provide a wealth of information on the
production process \cite{DEL06,DEL09,Per09,DEL08}.  The main
features of the anomalous soft photon phenomenon can be summarized as
follows:
\begin{enumerate}
\item
Anomalous soft photons are produced in association with hadron
production at high energies.  They are absent when there is no hadron
production \cite{DEL08}.
\item
The anomalous soft photon yield is proportional to the hadron yield.
\item
The transverse momenta of the anomalous soft photons are in the region
of many tens of MeV/c.
\item
The anomalous soft photon yield increases approximately linearly as
the number of neutral or charged produced particles increases, but,
the yield of anomalous soft photons increases much faster with
increasing neutral particle multiplicity than with charged particle
multiplicity.
\end{enumerate}

Previously, many different theoretical models have been put forth to
explain the anomalous soft photon phenomenon.  Reviews of the
experimental results and theoretical models have been presented
\cite{Bal90,Lic94}.  There are models based on the assumption of a
cold quark gluon plasma \cite{Van89}, boost-invariant classical flux
tube \cite{Czy94}, gluon dominance \cite{Kok07}, Unruh-Davies effect
\cite{Dar91}, synchrotron radiation in the stochastic nonperturbative
QCD vacuum \cite{Nac94}, the classical string fragmentation
\cite{And83}, closed quark-antiquark loop
\cite{Sim08}, and ADS/CFT Supersymmetric Yang-Mills
theory \cite{Hat10}.  These different proposed models are useful to
indicate that to search for the source of the anomalous soft photons,
it is necessary to go beyond the electromagnetic bremsstrahlung
process.  While the various models may explain some features of the
process, the fourth feature listed above from the recent DELPHI
observations cannot be explained by all existing models
\cite{DEL09,Per09}; a complete understanding of the basic
origin of the anomalous soft photon is still lacking.

We would like to propose a model for the simultaneous production
hadron and anomalous soft photons in the $q$-$\bar q$
string-fragmentation in quantum field theory to explain the anomalous
soft photon phenomenon.  As described by Casher, Kogut, and Suskind
\cite{Cas74}, the production of mesons in such a theory arises from
the oscillation of color charge densities of the quark vacuum in the
flux tube when a quark and an antiquark (or a diquark) pull away from
each other at high energies.  These color charge density oscillations
obey the Klein-Gordon equation characterized by the mass of the meson
\cite{Cas74,Bjo73,Bjo83,Sch62,Low71,Col75,Col76,Won91,Won94,Won09,Won09a}.
Because a quark carries both a color charge and an electric charge,
the underlying dynamical motion of the quarks in the vacuum that
generate color charge density oscillations will also generate electric
charge density oscillations in the flux tube.  These color charge
density oscillations will lead to the production of photons that are
clearly additional to those from the electromagnetic bremsstrahlung
process.  Thus the oscillation of the quark densities in the vacuum
will lead to the simultaneous color and electric charge density
oscillations and will subsequently lead to simultaneous and
proportional production of mesons and anomalous photons, in agreement
with the first two features of the anomalous soft photon phenomenon
listed in the beginning of this section.

It is of interest to examine whether the model also leads to results
that will be consistent with the remaining features of the anomalous
soft photon production phenomenon.  For such a purpose, we need to
know the properties of these electric charge density oscillations of
the quarks in the flux tube.  Will the electric charge density
oscillations also obey the Klein-Gordon equation characterized by a
mass to give rise to stable photons in the flux tube environment?  If
these photons are stable, what are the magnitudes of their masses and
how do the masses depend on the quantum numbers and other physical
properties?  If they are produced, how are they observed
experimentally in four-dimensional space-time?

To answer these questions, we start with quarks interacting with both
QCD4 and QED4 interactions in four-dimensional space-time in Section
II.  We specialize to the case of the flux tube formation for high
energy particle production processes under longitudinal dominance and
transverse confinement.  The system can then be approximately
compactified into a QCD2$\times$QED2 system in two-dimensional
space-time, with the coupling constants in different space-time
dimensions related by the flux tube radius. In Section, III, we
examine the non-Abelian bosonization of the QCD2$\times$QED2 system.
We find stable QCD2 and QED2 bosons for quarks with two flavors
arising from the density oscillations of the quarks in the flux tube.
These bosons can be identified as QCD2 mesons and QED2 photons.  The
boson masses are then expressed as a function of the coupling
constants and the quark transverse mass.  In Section IV, we estimate
the coupling constants and the boson masses for the case of the $Z^0$
hadronic decay.  As QCD2 mesons and QED2 photons are stable in the
flux tube environment, we can infer from the quantum field theory
description of the particle production process in Ref.\ \cite{Cas74}
that these QCD2 mesons and QED2 photons will be produced
simultaneously in $q$-$\bar q$ string fragmentation.  The present
model for the production of anomalous soft photons can also be
conveniently called the QED2 photon model.  In Section V, we discuss
the adiabatic decompactification of produced photons and mesons from
two-dimensional space-time to particles in four-dimensional
space-time.  In Section VI, we investigate how the QED2 model of
photon production may explain the experimental anomalous soft photon
transverse momentum distributions.  In Section VII, we examine the
rates of QED2 meson and QED2 photon production and the correlation of
the soft photon yield with charge and neutral particle multiplicities.
In Section VIII, we suggest future experimental measurements to test
the QED2 photon model.  In Section IX, we present our conclusions and
discussions.

\section{Flux Tube Environment in High Enegy Particle Production
  Processes}

We wish to investigate the process of soft photon production in
association with hadron production, in which the produced hadrons
consist mostly of mesons.  In the quantum field theory description of
meson production process as analogous to the particle production
process in quantum electrodynamics in two-dimensions (QED2), mesons
that are stable within the theory will be produced along the string,
when a quark and an antiquark at the two ends of the string pull apart
at high energies \cite{Cas74,Sch62,Low71,Col75,Col76} .  The rapidity
distribution of these produced mesons exhibits the property of boost
invariance in the limit of infinite energies \cite{Cas74,Bjo73,Bjo83}.
For a finite energy system, the boost-invariant solution turns
naturally into a rapidity plateau, whose width increases with energy
as $\ln(\sqrt{s})$ \cite{Won91,Won94}.

The $q$-$\bar q$ string is an idealization of a flux tube with a
transverse profile, which reveals itself as the transverse momentum
distribution of the produced particles \cite{Gat92}.  Experimentally,
the presence of a flux tube is evidenced by the limiting average
transverse momentum and a rapidity plateau
\cite{Cas74,And83,Bjo73,Bjo83,Won91,Won94,Won09,Won09a,Gat92} as in
high-energy $e^+$-$e^-$ annihilations
\cite{Aih88,Hof88,Pet88,Abe99,Abr99} and $pp$ collisions \cite{Yan08}.

To investigate the simultaneous production of QCD and QED quanta in
the flux tube fragmentation, we study quarks interacting with both QCD
and QED interactions, in circumstances leading to the formation of the
flux tube.  It becomes convenient to consider the U$(3)$ group which
breaks up into the color SU(3) and the electromagnetic U(1) subgroups.
The SU(3) and U(1) subgroups differ in their coupling constants and
communicative properties.  We introduce the generator $t^0$ for the
U$(1)$ subgroup,
\begin{eqnarray}
t^0=\frac{1}{\sqrt{6}}
\left ( \begin{matrix}  
                1 & 0 & 0 \cr
                0 & 1 & 0 \cr
                0 & 0 & 1 
        \end{matrix}   \right ),
\end{eqnarray}
which adds on to the eight generators of the SU$(3)$ subgroup,
$\{t^1,...t^8\}$, to form the nine generators of the U(3) group.  They
satisfy $ {\rm tr}\{ t^\alpha t^ \beta \} = \delta^{\alpha\beta}/2
{~~\rm for~~} \alpha,\beta=0,1,..,8$. 

Limiting our consideration to quarks with two light flavors, we
examine the QCD4$\times$QED4 system in four-dimensional space-time
$x^\mu$, with $\mu$=0,1,2,3.  The dynamical variables are the quark
fields, $\psi_f^a$, and the U(3) gauge fields, $A_\nu$=$A_\nu^\alpha
t^\alpha $, where $a$ is the color index with $a$=1,2,3, $f$ is the
flavor index with $f$=$u,d$, and $\alpha$ is the U(3) generator index
with $\alpha$=0,1,..,8.  The coupling constants $g_f^\alpha$ depend on
$\alpha$ and $f$ and are given explicitly by
\begin{eqnarray}
\label{qcdcc} 
g_u^{\{1,..,8\}}=g_d^{\{1,..,8\}}=g_{\rm QCD4}, {\rm~~for~~QCD},
\end{eqnarray}
\begin{eqnarray} 
\label{qedcc}
g_u^0=-e_u=-Q_u e_{\rm QED4}, ~~~g_d^0=-e_d=-Q_d e_{\rm QED4}   
{\rm~~for~~QED},
\end{eqnarray}
with $Q_u=2/3$, and $Q_d=-1/3$.  We use the convention of summation
over repeated indices, but the summation symbol and indices are
occasionally written out explicitly to avoid ambiguities.  For brevity
of notations, the indices $a$, $f$, and $\alpha$ in various quantities
are implicitly understood except when they are needed.  For example,
the $b$-color component of $(gA_\mu \psi)^b$ written explicitly is $
\sum_{f=u,d} \sum_{\alpha=0,...,8} \sum_{a=1,2,3} g_f^\alpha
A_\mu^\alpha (\tau^\alpha)^{ba} \psi_f^a$.

The transverse confinement of the flux tube can be represented by
quarks moving in a transverse scalar field $m({\bf r})$ where $m({\bf
  r})=S({\bf r})$+(current quark mass $m_q$) and $S({\bf r})$ is the
confining scalar interaction arising from nonperturbative QCD.  The
equation of motion of the quark field $\psi$ is
\begin{eqnarray}
\label{quark1}
\left \{ i \sD - m({\bf r}) \right \} \psi =0,
\end{eqnarray}
where 
\begin{eqnarray}
i\sD = \gamma^\mu \Pi_\mu =  \gamma^\mu ( p_\mu + g A_\mu).
\end{eqnarray}
The equation of motion for the gauge field $A_\mu$ is
\begin{eqnarray}
\label{Max4}
D_\mu F^{\mu \nu} = \partial_\mu F^{\mu \nu} -i g
[A_\mu, F^{\mu\nu}]
= g j^{\nu},  
\end{eqnarray}
where
\begin{eqnarray}
\label{F2}
F_{\mu \nu} = \partial_\mu A_\nu - \partial_\nu A_\mu   
-i g [A_\mu, A_\nu],
\end{eqnarray}
\begin{eqnarray}
\label{F3}
F_{\mu \nu}=F_{\mu \nu}^\alpha t^\alpha, 
~~~ j^\nu= j_{f\alpha}^{\nu} t^\alpha, 
\end{eqnarray} 
\begin{eqnarray}
\label{F4}
j_f^{\nu \,\alpha} =2 ~{\rm tr}~ {\bar \psi}_f \gamma^\nu t^\alpha \psi_f.
\end{eqnarray} 
Because of the commutative properties of the $t^0$ generator, the
commutator terms in Eqs.\ (\ref{Max4}) and (\ref{F2}) give zero
contributions for QED.  This set of equations describe the coupling of
the QCD and QED gauge fields and the quark fields, including those in
the quark vacuum.  The self-consistent coupling of these fields lead
to a problem of great complexity.  Fortunately, they can be simplified
under the conditions of longitudinal dominance and transverse
confinement that exist in the particle production environment at high
energies \cite{Won09a}.

This set of coupled equations in four dimensional space-time for
QCD4$\times$QED4 with the U(3) interaction has the same mathematical
structure as our previous set for QCD4 with the SU(3) interaction
\cite{Won09a}.  As the approximate compactification depends on the
separation of the transverse and the longitudinal degrees of freedom
and is independent of the nature of the underlying gauge group, there
should be similar approximate compactification of QED4$\times$QED4
into QCD2$\times$QED2.  We shall briefly summarize the salient points
leading to such an approximate compactification \cite{Won09a}.

In the problem of particle production at high energies leading to the
formation and fragmentation of the flux tube, we can focus our
attention on the self-consistent coupling of the quarks and the gauge
fields $A^\mu$ inside the tube.  In these high-energy processes, the
momentum scales for longitudinal dynamical motion of the leading $q$
and $\bar q$ as well as those of quarks in the underlying vacuum are
much greater than the momentum scales for their transverse motion such
that $|v^3|\gg |v^1|,|v^2|$, where ${\bf v}$ is a typical quark
velocity.  In the Lorentz gauge, the associated gauge field $A^\mu$ is
proportional to $(1,{\bf v})$. Under the dominance of the longitudinal
motion over the transverse motion in string fragmentation,
$|A^0|,|A^3| \gg |A^1|, |A^2|$.  Hence, inside the flux tube $A^1$ and
$A^2$ can be approximately neglected in comparison with the magnitudes
of $A^0$ or $A^3$.  It is further reasonable to assume that the gauge
fields $A^0$ and $A^3$ in the interior of the tube depend only weakly
on the transverse coordinates ${\bf r}=(x^1,x^2)$.  It is then
meaningful to investigate these fields inside the tube by averaging
them over the transverse profile of the flux tube.  After such an
averaging, $A^0$ and $A^3$ inside the tube can be considered as a
function of $(x^0,x^3)$ only.  As a consequence, the equation of
motion (\ref{quark1}) for the quarks becomes
\begin{eqnarray}
\label{quarkp}
\left \{ \gamma^0 \Pi_0 + \gamma^1 p_1 + \gamma^2 p_2 +
\gamma^3 \Pi_3 - m({\bf r}) \right \} \psi = 0.
\end{eqnarray}
We write the quark field $\psi(x)$ in terms of the longitudinal fields
$f_{\pm}$ and transverse fields $G_{1,2}$ with spinors $\xi_i$ as
\cite{Won91}
\begin{eqnarray}
\label{quark}
\psi(x)=[ G_1({\bf r})  \xi_1 - G_2({\bf r} \} \xi_2 ] f_+(x^0, x^3) 
+ [ G_1({\bf r}) \xi_3  +G_2({\bf r}) \xi_4 ] f_-(x^0, x^3) , 
\end{eqnarray}
where 
\begin{eqnarray}
\xi_1=\left ( \begin{matrix} 1\\
                     0\\
                     1\\
                     0
               \end{matrix}
      \right ),
~~~~
\xi_2=\left ( \begin{matrix} 0\\
                     1\\
                     0\\
                     -1
               \end{matrix}
      \right ),
~~~~
\xi_3=\left ( \begin{matrix} 1\\
                     0\\
                    -1\\
                     0
               \end{matrix}
      \right ),
{\rm ~~~and}~~~~
\xi_4=\left ( \begin{matrix} 0\\
                     1\\
                     0\\
                     1
               \end{matrix}
      \right ).
\end{eqnarray}
Working out the Dirac matrices in (\ref{quark1}), we obtain the
following set of coupled equations of motion 
\begin{subequations}
\begin{eqnarray}
\label{a}
[(\Pi^0-\Pi^3)(\Pi^0+\Pi^3)- m^2({\bf r})
-{\bf p}_T^2]G_1f_- &=& -[(p^1-ip^2)m({\bf r})]G_2f_- ,\\
\label{b}
[(\Pi^0-\Pi^3)(\Pi^0+\Pi^3)- m^2({\bf r})
-{\bf p}_T^2]G_2f_- &=& ~~[(p^1+ip^2)m({\bf r})]G_1f_- ,\\
\label{c}
[(\Pi^0+\Pi^3)(\Pi^0-\Pi^3)- m^2({\bf r})
-{\bf p}_T^2]G_1f_+ &=& -[(p^1-ip^2)m({\bf r})]G_2f_+ ,\\
\label{d}
[(\Pi^0+\Pi^3)(\Pi^0-\Pi^3)- m^2({\bf r})
-{\bf p}_T^2]G_2f_+ &=& ~~[(p^1+ip^2)m({\bf r})]G_1f_+ .
\end{eqnarray}
\end{subequations}
By the method of the separation of variables, we introduce the
eigenvalue (the quark transverse mass) $m_T$ for transverse motion,
\begin{subequations}
\begin{eqnarray}
\label{f}
[{\bf p}_T^2 + m^2({\bf r})-m_T^2]
G_1({\bf r}) &=&~~[(p^1-ip^2)m({\bf r})]G_2({\bf r}),\\
\label{ff}
[{\bf p}_T^2 + m^2({\bf r})-m_T^2]
G_2({\bf r}) &=&-[(p^1+ip^2)m({\bf r})]G_1({\bf r}),
\end{eqnarray}
\end{subequations}
and obtain the coupled equations for longitudinal motion,
\begin{subequations}
\begin{eqnarray}
\label{h}
[(\Pi^0-\Pi^3)(\Pi^0+\Pi^3)-m_T^2] f_-(x^0,x^3) &=&0,\\
\label{i}
[(\Pi^0+\Pi^3)(\Pi^0-\Pi^3)-m_T^2] f_+(x^0,x^3) &=&0.
\end{eqnarray}
\end{subequations}
If we introduce the two-dimensional Dirac spinor $\psi_{\rm 2D}$ as
\begin{eqnarray}
\label{quarkf}
\psi_{\rm 2D} = 
\left ( \begin{matrix} f_+\\
                       f_-
        \end{matrix}
      \right ),
\end{eqnarray}
and the 2-dimensional gamma matrices as \cite{Wit84,Abd96},
\begin{eqnarray}
\gamma_{\rm 2D}^0 = 
\left ( \begin{matrix} 0 & 1\\
                       1 & 0
        \end{matrix}
      \right ),
~~~~
\gamma_{\rm 2D}^3 = 
\left ( \begin{matrix} 0 & -1\\
                       1 & 0
        \end{matrix}
      \right ),
~~~~
\gamma_{\rm 2D}^0 \gamma_{\rm 2D}^3 = \gamma_{\rm 2D}^5=
\left ( \begin{matrix} 1 & 0\\
                       0 & -1
        \end{matrix}
      \right ),
\end{eqnarray}
then Eqs. (\ref{h}) and (\ref{i}) can be rewritten as the Dirac equation 
\begin{eqnarray}
\label{quark2d}
\left \{ \gamma_{\rm 2D}^0(p_0 + g A_0)+ \gamma_{\rm  2D}^3(p_3+gA_3)- m_T \right \} \psi_{\rm 2D} =0,
\end{eqnarray}
which is the equation of motion for a quark in two-dimensional gauge
fields of $A_0$ and $A_3$, except that the coupling constants $g$ are
those in four-dimensional space-time, $g_{\rm 4D}$, and the gauge
fields $A_{0(\rm 4D)}$ and $A_{3(\rm 4D)}$ are those determined from a
four-dimensional current source $j_{\rm 4D}^\nu$ given by Eqs.\
(\ref{F3}) and (\ref{F4}) involving four-dimensional quark fields
$\psi_{\rm 4D}$.  It is necessary to renormalize the coupling constants
and use quantities determined from two-dimensional source currents and
fields, to bring it to the proper two-dimensional space-time form.

Utilizing the result of Eq.\ (\ref{quark2d}) for the quark field and
using the quark wave function (\ref{quark}), the set of equations of
motion (\ref{quark1})-(\ref{F4}) along the longitudinal direction can
be cast into the forms of quarks and gauge fields interacting with the
QCD and QED in two-dimensional space time of $(x^0,x^3)$, by
transversely averaging Eq.\ (\ref{Max4}) over the profile of the flux
tube and by relating the coupling constants with the renormalization
\cite{Won09a}
\begin{eqnarray}
g_{\rm 2D}^2 = g_{\rm 4D}^2 \langle (|G_1({\bf r})|^2 + |G_2({\bf
r})|^2)\rangle_{{}_T}.
\end{eqnarray}
Such a renormalization yields a two-dimensional coupling constant
$g_{\rm 2D}$ that possesses the dimension of a mass.

After the coupling constant renormalization, the equations of motion
for the quark and the gauge fields in the longitudinal and time
directions are as given in the same form as Eqs.\
(\ref{quark1})-(\ref{F4}) in two-dimensional space-time
$x^\mu(\mu=0,3)$ with $\psi_{\rm 4D}$ replaced by $\psi_{\rm 2D}$,
$\gamma_{\rm 4D}^\mu$ gamma matrices replaced by the 2-dimensional
gamma matrices $\gamma_{\rm 2D}^\mu$, the quantity $m({\bf r})$
replaced by the quark transverse mass $m_T$, and gauge fields
$A_{\mu{\rm (4D)}}$ limited to $\mu=0$,3 and replaced by $A_{\mu({\rm
    2D})}$ determined from the Maxwell Equation (\ref{Max4}), $D_\mu
F_{\rm 2D}^{\mu \nu}=g_{\rm 2D} j_{\rm 2D}^\nu$, with two-dimensional currents
$j_{\rm 2D}^\nu$ that arise from $\psi_{\rm 2D}$.

We can get an approximate relation between $g_{\rm 2D}$ and $g_{\rm
4D}$ by considering the case of a uniform transverse flux tube profile
with a transverse radius $R_T$,
\begin{eqnarray}
(|G_1({\bf r})|^2 + |G_2({\bf
r})|^2) \sim \Theta(R_T - |{\bf r}|)/\pi R_T^2.
\end{eqnarray}
The coupling constants in two-dimensional space-time and 4-dimensional
space-time are then related approximately by \cite{Won09a}
\begin{eqnarray}
\label{est}
g_{\rm 2D}^2 \sim  \frac{g_{\rm 4D}^2}{\pi R_T^2}.
\end{eqnarray}
Such a result is expected from dimensional analysis, where the length
scale in going from a tube to a string involves only the flux tube
radius. The above relationship between the coupling constants in
different space-time dimensions will be used later to estimate the
boson masses.

\section{Bosonization of QCD2$\times$QED2 for quarks with two flavors}

Under the longitudinal dominance and transverse confinement, the
QCD4$\times$QED4 system can be approximate compactified as the
QCD2$\times$QED2 system with a quark transverse mass $m_T$.  The flux
tube becomes the arena for the quarks in the underlying vacuum to
interact self-consistently with the QCD and QED gauge fields.  We
shall henceforth work with QCD2$\times$QED2 in two-dimensional
space-time.  For brevity of notation, the two-dimensional designation
of various quantities will be understood in what follows. The
Lagrangian density for QCD2$\times$QED2 that corresponds to the
two-dimensional version of Eqs.\ (\ref{quark1})-(\ref{F4}) is
\begin{eqnarray}
\label{lag}
{\cal L}= {\bar \psi} [ \gamma^\mu(i\partial_\mu +g A_{\mu}) - m_T ]
\psi - \frac{1}{4} F_{\mu \nu} F^{\mu \nu}. 
\end{eqnarray}
As in the Section II, the color index $a$, the flavor index $f$, and
the U(3) generator index $\alpha$ are implicitly understood, and the
summation convention is used.

We wish to search for bound states arising from the density
oscillations of the color and electric charges of the quarks in
QCD2$\times$QED2 in the strong coupling limit, in which the strength of
the QCD2 interaction is much greater than the quark mass.  The best
method to search for bound states is by bosonization in which bosons
are bound and nearly free, with residual sine-Gordon interactions that
depend on the quark mass \cite{Col76},\cite{Man75}-\cite{Nag09}.

The U(3) gauge interactions under consideration contains the
non-Abelian color SU(3) interactions.  Consequently the bosonization
of the color degrees of freedom should be carried out according to the
method of non-Abelian bosonization which preserves the gauge group
symmetry \cite{Wit84}.

While we use non-Abelian bosonization for the U(3) interactions, we
shall use the Abelian bosonization for the flavor degrees of freedom.
This involves keeping the flavor labels in the bosonization without
using the flavor group symmetry.  Although the Abelian bosonization in
the flavor sector obscures the isospin symmetry in QCD, the QCD
isospin symmetry is still present.  It can be recovered by complicated
non-linear general isospin transformations \cite{Col76,Hal75}.

As in any method of bosonization, the non-Abelian method will succeed
for systems that contain stable and bound boson states with relatively
weak residual interactions.  Thus, not all the degrees of freedom
available to the bosonization technique will lead to good boson states
with these desirable properties. For example, some of
the bosonization degrees of freedom in color SU(3) may correspond to
bosonic excitations into colored objects of two-fermion complexes and
may not give rise to stable bosons.  It is important to judiciously
search for those boson degrees of freedom that will eventually lead to
stable and bound bosons.

Keeping this perspective in our mind, we can examine the non-Abelian
bosonization of the system under the U(3) interactions.  The
non-Abelian bosonization program consists of introducing boson fields
to describe an element $u$ of the U(3) group and showing subsequently
that these boson fields lead to stable bosons with finite or zero
masses.

In the non-Abelian bosonization, the current $j_\pm$ in the light-cone
coordinates, $x^{\pm}$=$(x^0 \pm x^3)/\sqrt{2}$, is bosonized as
\cite{Wit84}
\begin{subequations}
\label{jj}
\begin{eqnarray}
\label{jja}
j_+ & = & ~~(i/2\pi) u^{-1} (\partial_+ u),\\
\label{jjb}
j_- & = & -(i/2\pi) (\partial_- u) u^{-1}.
\end{eqnarray}
\end{subequations}
An element of the U(1) subgroup of the U(3) group can be represented
by the boson field $\phi^0$
\begin{eqnarray}
u = \exp\{ i 2 \sqrt{\pi} \phi^0 t^0 \}.
\end{eqnarray}  
Such a bosonization poses no problem as it is an Abelian subgroup.  It
will lead to a stable boson as in Schwinger's QED2.

To carry out the bosonization of the color SU(3) subgroup, we need to
introduce boson fields to describe an element $u$ of SU(3).  There are
eight $t^\alpha$ generators which provides eight degrees of freedom.
We may naively think that for the non-Abelian bosonization of SU(3),
we should introduce eight boson fields $\phi^\alpha$ to describe $u$
by
\begin{eqnarray}
u = \exp\{ i 2 \sqrt{\pi} \sum_{\alpha=1}^8 \phi^\alpha t^\alpha\}.
\end{eqnarray}
However, a general variation of the element $\delta u /\delta x^\pm$
will lead to quantities that in general do not commute with $u$ and
$u^{-1}$, resulting in $j_\pm$ currents in Eqs.\ ({\ref{jj}) that are
complicated non-linear admixtures of the boson fields $\phi^\alpha$.
It will be difficult to look for stable boson states with these
currents.

We can guide us to a situation that has a greater chance of finding 
stable bosons by examining the bosonization problem from a different
viewpoint.  We can pick a unit generator ${\bf n}=\{n^1,n^2,..,n^8\}$ oriented
in any direction of the eight-dimensional $\alpha$-space and can
describe an SU(3) group element $u$ by an amplitude $\phi$ and the
unit vector $\bf n$,
\begin{eqnarray}
\label{choi}
u = \exp\{ i 2 \sqrt{\pi} \phi \sum_{\alpha=1}^8 n^\alpha t^\alpha\}.
\end{eqnarray}
The boson field $\phi$ describes one degree of freedom, and the
direction cosines $\{n^\alpha,\alpha=1,..,8\}$ of the unit vector $\bf
n$ describe the other seven degrees of freedom.  A variation of the
amplitude $\phi$ in $u$ while keeping the unit vector orientation
fixed will lead to a variation of $\delta u/\delta x^\pm $ that will
commute with $u$ and $u^{-1}$ in the bosonization formula (\ref{jj}),
as in the case with an Abelian group element.  It will lead to simple
currents and stable QCD2 bosons with well defined masses, which will
need to be consistent with experimental QCD meson data.  On the other
hand, a variation of $\delta u/\delta x^\pm $ in any of the other
seven orientation angles of the unit vector $\bf n$ will lead to
$\delta u/\delta x^\pm $ quantities along other $t^\alpha$ directions.
These variations of $\delta u/\delta x^\pm $ will not in general
commute with $u$ or $u^{-1}$.  They will lead to $j_\pm$ currents that
are complicated non-linear functions of the eight degrees of
freedom. We are therefore well advised to search for stable bosons by
varying only the amplitude of the $\phi$ field, keeping the
orientation of the unit vector fixed, and forgoing the other seven
orientation degrees of freedom.

As a unit vector $\bf n$ in any orientation can be rotated to the
first axis along the $t^1$ direction by an orthogonal transformation
in the $\alpha$-space, we can consider the unit vector $\bf n$ to lie
along the $t^1$ direction without a loss of generality.  For the U(3)
group, The appropriate bosonization program that will eventually lead
to stable bosons is to limit the consideration to only the $\phi^0$
and $\phi^1$ degrees of freedom.  We are therefore justified to
bosonize an element $u$ of the U(3) group as
\begin{eqnarray}
\label{uua}
u = \exp\{ i 2 \sqrt{\pi} \sum_{\alpha=0}^1 \phi^\alpha t^\alpha\}.
\end{eqnarray}
From Eqs.\ (\ref{jja}) and  (\ref{jjb}), we obtain then
\begin{eqnarray}
j_{f\pm} = \mp \frac{1} {\sqrt{\pi}} \sum_{\alpha=0}^1 
(\partial_\pm \phi_f^\alpha) t^\alpha,
\end{eqnarray}
where we have written out the flavor index explicitly.  The gauge
fields can be easily obtained by using the $A_-=0$ gauge for which
terms involving the commutators in Eqs.\ (\ref{Max4}) and (\ref{F2})
vanish.  The Maxwell equation becomes
\begin{eqnarray}
-\partial_-^2 A_+ = - gj^+ 
\end{eqnarray}
and the solution is 
\begin{eqnarray}
A_+ =   \frac{g}{ \partial_-^2} j^+. 
\end{eqnarray}
The interaction energy becomes
\begin{eqnarray}
H_I = \frac{1}{2} \int dx^- j_f^+ \frac{g}{\partial_-^2} j_f^+ 
    = \frac{1}{4\pi} \int dx^-\sum_{\alpha=0}^1 
      (\sum_{f=u,d} g_f^\alpha \phi_f^\alpha)^2 .
\end{eqnarray}
The kinetic energy term of the Lagrangian density, $ {\bar \psi}
\gamma^\mu ~i\partial_\mu \psi$ , can be bosonized as \cite{Wit84}
\begin{eqnarray}
{\cal L}_{\rm KE}= 
\frac{1}{8\pi} \sum_{f=u,d} 
{\rm tr} ~ \left ( \partial _\mu u_f  \partial^\mu u_f^{-1} \right ),
\end{eqnarray}
as the Wess-Zumino term for $u$ in the form of Eq.\ (\ref{uua}) gives
no contribution. Eq.\ (\ref{uua}) then leads to
\begin{eqnarray}
{\cal L}_{\rm KE}= 
\frac{1}{4} \sum_{f=u,d} \sum_{\alpha=0}^1 
 \partial _\mu \phi_f^\alpha  \partial^\mu \phi_f^\alpha .
\end{eqnarray}
The mass term involves the scalar density $\bar \psi \psi$ which can
be bosonized as
\begin{eqnarray}
\label{psipsi}
:\bar \psi \psi : &\to& - \frac{e^\gamma}{2\pi} \mu N_{\mu} 
\sum_{f=u,d} {\rm tr}~\left (\frac{u_f + u_f^{-1}}{2} \right ) \nonumber \\
&=&  - \frac{e^\gamma}{2\pi} \mu N_{\mu}   
\sum_{f=u,d}
{\rm tr}
[\cos(2 \sqrt{\pi}\sum_{\alpha=0}^1 \phi_f^\alpha t^\alpha)],
\end{eqnarray}
where $\gamma=0.5772$ is the Euler constant, $N_\mu$ is
normal ordering with respect to the mass scale $\mu$ for the problem
in question.  It is easy to show that
\begin{eqnarray}
{\rm tr}
[\cos(2 \sqrt{\pi}\sum_{\alpha=0}^1 \phi_f^\alpha t^\alpha)]
= 2 \cos (2 \sqrt{\pi/6} \phi_f^0) \cos (2 \sqrt{\pi/4} \phi_f^1).
\end{eqnarray}
We shall not examine the zero mode and the theta vacuum in the present
exploratory study.  We obtain the Hamiltonian density
\begin{eqnarray}
\label{hh}
{\cal H}&=& \frac {1}{2}N_\mu \sum_{\alpha=0}^1 \biggl \{ \sum_{f=u,d}
 \left [ \frac{1}{2} (\Pi_f^\alpha)^2 + \frac{1}{2}
 (\partial_1\phi_f^\alpha )^2 \right ] + \frac{1}{ 2\pi} (\sum_{f=u,d}
 g_f^\alpha \phi_f^\alpha)^2 \biggr \} \nonumber\\ 
& & - \frac{e^\gamma m_T \mu}{2\pi}2 N_\mu \sum_{f=u,d} 
 \cos(2\sqrt{\pi/6}\phi_f^0)  \cos(2\sqrt{\pi/4}\phi_f^1).
\end{eqnarray}

In the flavor sector, the up quark has isospin quantum numbers
$(I,I_3)$=$(1/2,1/2)$ and the down quark has $(I,I_3)$=$(1/2,-1/2)$.
The up and down quarks combine to form the isoscalar $(I,I_3)$=$(0,0)$
state and the isovector $I$=$1$ states, which split into three
components with $I_3$=$(1,0,-1)$.  Because the quark electric charge
$Q_f$ depends on the flavor quantum number, there is no isospin
symmetry for QED2, and the four states split apart.  We shall focus
our attention only on the isoscalar $(I,I_3)$=$(0,0)$ QED2 state and
the isovector $(I,I_3)$=$(1,0)$ QED2 state.  The other two
$(I,I_3)$=$(1,\pm1)$ QED2 states involve composite constituents with
like electric charges and are unlikely to be stable in the
electromagnetic sector.  For brevity of nomenclature, we shall refer
to the isovector $(I,I_3)$=$(1,0)$ photon simply as isovector photon,
with the qualifying specification `$I_3$=$0$' implicitly understood.

In QCD with two flavors, the isospin symmetry remains a good symmetry,
which is weakly broken by a small difference between the up and down
quark masses. Thus, the QCD quark-antiquark meson states are
specified by isospin quantum numbers $I$ with nearly degenerate $2I+1$
members of different $I_3$ components.  The knowledge of the location
of the $(I,I_3)$=$(1,0)$ QCD state allows us to infer the locations of
the other two QCD $(I,I_3)$=$(1,\pm 1)$ states.

We can construct the $\phi_I^\alpha$ fields for the isospin
$(I,I_3=0)$ states, for up and down quark fields moving in phase or
out of phase,
\begin{eqnarray}
\phi_I^\alpha = \frac{1}{\sqrt{2} } \left [ \phi_u^\alpha +(-1)^I
\phi_d^\alpha \right ].
\end{eqnarray}
We can also construct the corresponding isospin canonical momenta
\begin{eqnarray}
\Pi_I^\alpha = \frac{1}{\sqrt{2} }\left [ \Pi_u^\alpha +(-1)^I 
\Pi_d^\alpha \right ].
\end{eqnarray}
The Hamiltonian density in terms of boson fields of different
isospin quantum numbers $I$ and the same $I_3=0$ is
\begin{eqnarray}
\label{ham0}
{\cal H}= \frac{1}{2}   N_\mu \biggl \{ 
   \sum_{\alpha=0}^1 
\sum_{I=0}^1 \biggl [\frac{1}{2}  (\Pi_I^\alpha)^2 
               + \frac{1}{2}  (\partial_1 \phi_I^\alpha)^2        
\biggr ] +  V (\{\phi_I^\alpha\}) 
                              \biggr \},
\end{eqnarray}
where $ V (\{\phi_I^\alpha\})= V_{\rm int}(\{\phi_I^\alpha\}) + V_{\rm
m}(\{\phi_I^\alpha\})$ with the interaction energy
\begin{eqnarray}
 V_{\rm int} (\{\phi_I^\alpha\}) 
= 
\frac{1}{2}\left ( \sum_{I=0}^1 \frac{g_u^\alpha + (-1)^I g_d^\alpha}
                    {\sqrt{2}\pi} \phi_0^\alpha \right )^2,        
\end{eqnarray}
and the quark mass term 
\begin{eqnarray}
 V_{\rm m} (\{\phi_I^\alpha\}) 
=- \frac{e^\gamma m_T \mu}{2\pi} 2 \left [ \prod_{I=0}^1
\cos\left (\sqrt{2\pi}(\frac{\phi_I^0}{\sqrt{6}} +\frac{\phi_I^1}{\sqrt{4}})
\right )
+ \prod_{I=0}^1 \cos \left (\sqrt{2\pi}(\frac{\phi_I^0}{\sqrt{6}}
-\frac{\phi_I^1}{\sqrt{4}}) \right ) \right ].
\end{eqnarray}
We can get the gross features of the system by expanding the potential
about the minimum located at $\phi_0^\alpha=0$ and $\phi_1^\alpha=0$.
Evaluating the second derivatives of the potential at the potential
minimum, we obtain the mass square $(M_{I({\rm 2D})}^\alpha)^2$ of
stable boson quanta for $\alpha$=0,1,
\begin{eqnarray}
\label{pot}
(M_{I({\rm 2D})} ^\alpha)^2
=\biggl [ \frac{\partial^2}{\partial (\phi_{I}^\alpha)^2} 
V(\{\phi_I^\alpha\})\biggr ]_{\phi_0^\alpha,\phi_1^\alpha=0} 
=\left ( \frac{g_u^\alpha+(-1)^I g_d^\alpha}{\sqrt{2\pi}} \right )^2
+ \frac{2}{3-\alpha} e^\gamma m_T \mu. 
\end{eqnarray}
The Hamiltonian density (\ref{ham0}) represents a QCD2 and QED2 system
of isoscalar and isovector boson fields $\phi_I^\alpha$ whose field
quanta acquire the mass $M_{I({\rm 2D})}^\alpha$, where $\alpha=0$ for
QED2 and $\alpha=1$ for QCD2.  As the boson field $\phi_I^\alpha$ is
related to the gauge field $A_+$ through Eqs.\ (29) and (31), the
quanta of $\phi_I^\alpha$ are also the quanta of the gauge fields
$A^+$.  The QCD2 bosons and QED2 bosons can be appropriately called
QCD2 mesons and QED2 photons respectively.

Because the righthand side of Eq.\ (\ref{pot}) is a non-negative
quantity with $(M_{I({\rm 2D})}^\alpha)^2\ge 0$, these QCD2 mesons and
QED2 photons are stable bosons.  They acquires a mass because a gauge
field oscillation leads to a quark density oscillation, and through
the Maxwell equation the quark density oscillation in turn leads to a
gauge field oscillation, which in turn modifies the quark density
oscillation.  The self-consistency of gauge field oscillations and the
induced quark density oscillations lead to an equation of motion for
the gauge field oscillation in the form of a Klein-Gordon equation
characterized by a mass.

Our result of the boson masses in Eq.\ (\ref{pot}) represents a
QCD2$\times$QED2 generalization of previous results in
\cite{Col76},\cite{Hal75}-\cite{noteH},
 where QED2 and QCD2 have been
examined separately.  In the massless quark limit, the QCD2 and QED2
boson masses are given by $|g_u^\alpha + (-1)^I
g_d^\alpha|/\sqrt{2\pi}$.  In this limit, the QCD2 masses are the same
as what one obtains by using QED2 and replacing the electric charges
in QED2 with the color charges in QCD as in \cite{Col76}.  This
equivalence of the Abelian QED2 solution and the non-Abelian QCD2
solution in the massless limit arises because our judicious search for
stable QCD2 bosons in the non-Abelian bosonization of SU(3) requires
the variation of only the amplitude $\phi$ while the orientation of
${\bf n}$ in Eq.\ (\ref{choi}) is held fixed.  The non-Abelian
bosonization in QCD2 that results in stable QCD mesons is in effect
Abelian in nature.  This explains why previous Abelian QED2 results of
boson masses \cite{Col76} and string fragmentation \cite{Cas74} can be
applied to the non-Abelian QCD problems by replacing the electric
charges in QED2 with the color charges in QCD.

Previously, Abelian-type solutions were obtained for multiflavor QCD2
mesons using non-Abelian bosonization for both the color and flavor
degrees of freedom in the large $N_f$ limit \cite{Arm99}.  The mass of
the single massive boson in the massless quark limit was found to be
$M_{({\rm QCD2})}$=$g_{_{QCD2}}\sqrt{N_f/\pi}$
\cite{Arm99,Tri02,Abr04,Abr05,notemass}.  Our QCD2$\times$QED2
analysis here indicates that Abelian-type solutions exist also for
QCD2 mesons for quarks with two flavors, and is not limited to the
large $N_f$ limit.  Our mass of the QCD2 isoscalar meson in the
massless quark limit is $M_{0({\rm
    QCD2})}^1$=$g_{_{QCD2}}\sqrt{2/\pi}$, which matches the mass of
the massive boson of \cite{Arm99,Tri02,Abr04,Abr05,notemass} for
$N_f$=$2$.  Thus, by using the non-Abelian bosonization in QCD2 but Abelian
bosonization in the flavor degrees of freedom in the present
treatment, the solutions of \cite{Arm99,Tri02,Abr04,Abr05} in the
large $N_f$ limit can be extended down to $N_f$=$2$.

In the massless quark limit (for $m_T \mu $=0 in this case), the
QCD2$\times$QED2 bosons are free.  With a finite value of $m_T \mu$,
they interact with a sine-Gordon residual interaction whose
strength depends on $m_T \mu$.  The present treatment places the QED2
mesons and the QCD2 photons on a parallel footing and allows the mutual
interaction between QCD2 mesons and QED2 photons.  To exhibit the
mutual interaction, it is instructive to expand the quark mass term in
powers of $\phi_I^\alpha$.  Up to the fourth order in $\phi_I^\alpha$,
we obtain
\begin{eqnarray}
V_{\rm m}(\{\phi_I^\alpha\})= 
\frac{1}{2} \sum_{\alpha=0}^1 \sum_{I=0}^1
      a_I^\alpha (\phi_I^\alpha)^2 
+\frac{1}{4} \sum_{\alpha=0}^1 \sum_{I,I'=0}^1
        b_{II'}^\alpha (\phi_I^\alpha)^2 (\phi_{I'}^\alpha)^2 
+\frac{1}{4} \sum_{\alpha=0}^1 \sum_{I=0}^1
           c_{I I'}  (\phi_I^0)^2 (\phi_I^1)^2
\end{eqnarray}
where
\begin{subequations}
\begin{eqnarray}
a_I^\alpha     &=& ~~\frac{2}{3-\alpha} e^\gamma m_T \mu,\\
b_{II'}^\alpha &=& -\frac{2\pi} {(3-\alpha)^2} e^\gamma m_T \mu,\\
c_{I I'}       &=& -\frac{\pi}{3}  e^\gamma m_T \mu.
\end{eqnarray}
\end{subequations}
Here the $a_I^\alpha$ coefficients give the contribution to
$(M_{I({\rm 2D})}^\alpha)^2$ from the quark mass term in Eq.\
(\ref{pot}).  The $b_{II'}^\alpha$ coefficients give the interaction
between bosons of the same type $\alpha$, and $c_{I I'}$ give the
interaction between QCD2 mesons and QED2 photons.  The negative signs
of the $b$ and $c$ coefficients indicate that the interaction between
the bosons are attractive in nature.

Previously, Coleman obtained the correction to the QED2 boson masses
arising from a non-zero quark mass, using the method of
renormal-ordering.  The mass correction was also obtained by examining
QED2 on a circle \cite{Hos98}, near-light-cone coordinates
\cite{Var96}, and mass-perturbation theory \cite{Ada97}.  We shall not
consider these refinements and contend ourselves with the estimate
using the second derivatives of the potential $V(\{\phi_I^\alpha\})$
in the present exploratory study.

\section{ QCD2 Meson and QED2 Photon Masses for Quarks with Two Flavors}

We consider first the boson masses in the massless quark limit because
they represent well-defined references.  In this limit, the boson
masses depend only on the coupling constants which acquire the
dimension of a mass as a result of the compactification.  They depend
on the flux tube radius as given by Eq.\ (\ref{est})
\cite{Won09,Won09a}.  For QCD in the flux tube, the QCD2 coupling
constant is given by
\begin{eqnarray}
\label{R1}
g_{\rm QCD2}^2 \sim 
\frac{g_{\rm QCD4}^2} {\pi R_T^2}
=\frac{g_{\rm QCD4}^2} {4\pi}\frac{4}{R_T^2}
=\frac{4\alpha_s}{R_T^2},
\end{eqnarray}
where $\alpha_s=g_{\rm QCD4}^2/4\pi$ is the strong interaction
coupling constant.
Similarly, for QED2 in the flux tube, the QED2 coupling constant is
given by
\begin{eqnarray}
\label{e1}
e_{\rm QED2}^2 \sim 
\frac{e_{\rm QED4}^2} {\pi R_T^2}
=\frac{e_{\rm QED4}^2} {4\pi}\frac{4}{R_T^2}
=\frac{4\alpha}{R_T^2},
\end{eqnarray}
where $\alpha=e_{\rm QED4}^2/4\pi=1/137$ is the fine structure
constant.  The magnitude of the flux tube radius $R_T$ is revealed by
the root-mean-squared transverse momentum of produced hadrons (mostly
pions) as
\begin{eqnarray}
\label{R2}
R_T \sim \frac{1}{\sqrt{\langle p_T^2\rangle_\pi } },
\end{eqnarray}
which empirically is slightly energy-dependent \cite{Won09a}.  We
shall focus our attention on the case of particle production in high
energy $e^+$-$e^-$ annihilations in the hadronic decay of $Z^0$.  The
measurement of the $\pi^0$ spectra in $Z^0$ hadronic decay gives
$\sqrt{\langle p_T^2\rangle_\pi }=0.56$ GeV in the reaction plane
\cite{Bar97} and thus the flux tube has a radius $R_T \sim 0.35$ fm.
For the strong coupling constant at this energy, we shall take
$\alpha_s=0.316$, which leads from Eq.\ (\ref{R1}) to the string
tension coefficient \cite{Won09,Won09a}
\begin{eqnarray}
\label{qcd2}
b={g_{\rm QCD2}^2}/{2}=0.2 {\rm ~GeV}^2,
\end{eqnarray}
and
\begin{eqnarray}
g_{\rm QCD2} =   0.632 {\rm ~GeV}.
\end{eqnarray}
From Eq.\ (\ref{e1}), the QED2 electromagnetic coupling constant has
the value
\begin{eqnarray}
\label{ee}
e_{\rm QED2}  \sim  0.096 {\rm ~GeV}.
\end{eqnarray}

With these coupling constants ($g_u^1$=$g_d^1$=$g_{\rm QCD2}$,
$g_u^0$=$-Q_u e_{\rm QED2}$, and $g_d^0$=$-Q_d e_{\rm QED2})$, the
values of QCD2 and QED2 boson masses in the massless quark limit are
shown in Table I.  One observes that QCD2 for quarks with two flavors
gives a massless pion in the massless quark limit, in agreement with
the concept of the pion being a Goldstone boson in the standard QCD
theory.  The isovector QCD2 meson lies lower than the isoscalar QCD2
meson at 504 MeV, whereas the ordering is opposite for the QED2
photons, with an isoscalar QED2 photon at 12.8 MeV and an isovector
QED2 photon at 38.4 MeV.  These QED2 photons lie in the region of
observed anomalous soft photons.
\begin{table}[h]
  \caption { QED2 and QCD2 boson masses obtained with $R_T$=0.35 fm
and  $g_{\rm QCD2}^2$=$2b$=0.4 GeV$^2$.  }
\vspace*{0.2cm} 
\hspace*{-0.0cm}
\begin{tabular}{|c|c|c|c|}
\cline{3-4} \multicolumn{2}{c|}{} & QCD2 & QED2 
       \\  
            \multicolumn{2}{c|}{} &      & 
       \\ \hline 
\multicolumn{2} {|c|} {\hspace*{-0.8cm}Coupling Constant} 
& $g_{\rm QCD2}$=632.5 MeV & $e_{\rm QED2}$=96 MeV         \\ \cline{1-4} 
massless quarks  & isoscalar boson mass  $M_{0({\rm 2D})}$ 
&  504.6 MeV &  12.8 MeV 
 \\  
        & isovector boson mass $M_{1({\rm 2D})}$
&  0         &  38.4 MeV 
 \\ \cline{1-4}
$m_T$=400 MeV & isoscalar boson mass $M_{0({\rm 2D})}$
&  734.6 MeV &  
        \\  
$\mu$=$m_T$~~~~~   & isovector boson mass $M_{1({\rm 2D})}$   
&  533.8 MeV &  
           \\ \cline{1-4}  
 $m_T$= 400 MeV  & isoscalar boson mass $M_{0({\rm 2D})}$
&            &  $O$(25.3 MeV) 
 \\  
$\mu$=$m_q$=$O$(1 MeV) & isovector boson mass $M_{1({\rm 2D})}$  
&                 &  $O$(44.1 MeV) 
  \\ \cline{1-4}  
\end{tabular}
\end{table}

Equation (\ref{pot}) indicates that the boson masses depend on four
mass scales: $g_{\rm QCD2}$, $e_{\rm QED2}$, $m_T$, and $\mu$.  In
addition to the coupling constants we have just discussed, we need to
specify the values of the transverse mass $m_T$ and the mass scale
$\mu$.  The discussions in the Section II indicate that as quarks
resides in the flux tube environment, they acquire a transverse mass
$m_T$.  The presence of the factor $m_T$ in Eq.\ (\ref{pot}) takes into
account the effects of non-perturbative chiral symmetry breaking and
transverse confinement that lead to the formation of the flux tube.
Because a pion is a quark-antiquark composite, we can estimate the
quark transverse mass $m_T$ from the pion transverse momentum, $m_T
\sim \sqrt{\langle p_T^2\rangle_\pi /2}$. For $Z^0$ hadronic decay,
$\sqrt{\langle p_T^2\rangle_\pi}=0.56$ GeV and we have $m_T \sim 0.4$
GeV.

The boson masses depend also on the mass scale $\mu$, which arises
from the bosonization of the scalar density ${\bar \psi} \psi$ as
given in Eq.\ (\ref{psipsi}).  The scalar density ${\bar \psi} \psi$
diverges in perturbation theory and has to be renormalized such that
$\langle {\bar \psi} \psi \rangle$=0 in a free theory.  It will need
to be renormal-ordered again in an interacting theory \cite{Col76}.
The scalar density and the corresponding mass scale therefore depends
on the interaction.  The dependence of the scalar density on the
interaction is also evidenced by the fact that the scalar density
${\bar \psi} \psi$ in Eq.\ (\ref{psipsi}) can be expanded in terms of
$ t^0$ and $t^1$, each of which is the generator of a different
interaction associated with a different coupling constant.  For the
strong interaction of QCD, confinement and chiral symmetry breaking
dominate and lead to a transverse mass $m_T$ that is much greater than
the current quark mass.  It is reasonable to take the mass scale $\mu$
in QCD to be the same as the quark transverse mass $m_T$
characterizing the flux tube transverse confinement and the presence
of chiral symmetry breaking.  Meson masses in QCD2 calculated with the
mass scale $\mu=m_T=0.4$ GeV is given in Table I.  It gives a QCD2
isovector meson mass of 0.534 GeV and a QCD2 isoscalar meson mass of
0.735 GeV in the flux tube.

For a theory with a relatively weak interaction such as QED, the
scalar density ${\bar \psi} \psi$ that diverges in perturbation theory
has to be renormalized in a nearly-free field in which the quark
energy is just the current quark masses.  The mass scale $\mu$ for
QED2 should therefore be the QED current quark mass $m_q$ appropriate
for a nearly-free theory.  The current quark mass $m_q$ associated
with perturbative QCD has the value of 1.5-6 MeV \cite{PDG08}. The
current quark mass associated with perturbative QED is not known and
presumably is of the same order of an MeV.  For lack of a more
definitive determination, we shall take $\mu=1$ MeV to calculate the
orders of magnitude of the QED2 photon masses.  The values of the QED2
boson masses obtained with $\mu=1$ MeV are given in Table I, which
gives an isoscalar photon of order 25 MeV and an isovector photon of
order 44 MeV.  They fall within the same order of magnitude of the
transverse momenta of anomalous soft photons.

\section {Adiabatic Decompactification of Bosons from Two-Dimensional 
to Four-Dimensional Space-time}

We have thus found that in the system of quarks with two flavors, the
boson quanta of QCD2 and QED2 are stable with masses that depend on
the isospin quantum numbers.  We can therefore infer from the quantum
field theory description of particle production in Ref.\ \cite{Cas74}
that these QCD2 mesons and QED2 photons will be produced
simultaneously in the same process of $q$-$\bar q$ string
fragmentation, when a quark pulls away from an interacting antiquark
at high energies.  

After a particle is produced in two-dimensional space-time how does it
decompactify in the four-dimensional space-time?  An appropriate way
to describe the decompactification is to identify the mass of the
particle in the two-dimensional theory as the transverse mass of the
particle in four-dimensional space-time.  This clearly works in the
case of the quark. In reverting back into the four-dimensional
space-time, the `quark mass' of $m_T$ in two-dimensional space-time
reverts back into the transverse mass of the quark in four-dimensional
space-time.  

After a boson is produced and the system expands longitudinally, the
interaction between the produced bosons weakens.  The produced boson
will subsequently emerge from the production region out to the
non-interacting region and will obey the mass shell condition.
We can consider a produced boson of mass
$M_{I({\rm 2D})}^\alpha$ of isospin $I$ and type $\alpha$ in
two-dimensional space-time.  The kinematic variables of the boson are
$E_{({\rm 2D})}$ and $p_{z({\rm 2D})}$, which obey the mass shell
condition $E_{({\rm 2D})}^2$=$p_{z({\rm 2D})}^2$+$(M_{I({\rm
2D})}^\alpha)^2$.  In the four-dimensional space-time, kinematic
variables of this particle are $E_{({\rm 4D})}$, $p_{z({\rm 4D})}$,
and ${\bf p}_{T({\rm 4D})}$.  The kinematic variables satisfy the mass
shell condition $E_{({\rm 4D})}^2$=$p_{z({\rm 4D})}^2$+${\bf
p}_{T({\rm 4D})}^2$+$(M_{I({\rm 4D})}^\alpha)^2 $, where $M_{I({\rm
4D})}^\alpha$ is the rest mass of the particle in four-dimensional
space-time.

In the emergence of the boson from two-dimensional space time to
four-dimensional space time, we envisage an adiabatic transverse
expansion from the two-dimensional flux tube to four-dimensional
space-time.  The adiabatic transverse expansion involves no change of
the particle energy and particle longitudinal momentum so that
$E_{({\rm 2D})}$=$E_{({\rm 4D})}$ and $p_{z({\rm 2D})}$=$p_{z({\rm
4D})}$.  Consequently, the mass shell conditions of the boson give
$(M_{I({\rm 2D})}^\alpha)^2$=${\bf p}_{T({\rm 4D})}^2$+$(M_{I({\rm
4D})}^\alpha)^2$=$(M_{IT}^\alpha)^2$, with the boson mass $M_{I({\rm
2D})}^\alpha$ in two-dimensional space-time turning into the boson
transverse mass $ M_{IT}^\alpha$ in four-dimensional space-time.

We can test the consistency of such a correspondence for the
production of mesons.  We consider the production of an isovector
meson, which is a pion.  Experimentally, a pion is produced with an
average $\sqrt{\langle p_T^2 \rangle}_\pi \sim 0.56$ GeV \cite{Bar97}
for the $Z^0$ hadronic decay.  Thus, the experimental (average)
isovector meson (pion) transverse mass is
\begin{eqnarray}
\label{pionmt}
M_{1T}^h=\sqrt{(0.14)^2+(0.56)^2} ~{\rm GeV} = 0.579 {\rm ~ GeV}.
\end{eqnarray}
We consider next the production of an isoscalar meson, which is the
$\eta$ meson with a rest mass $M_\eta=0.547$ GeV.  Experimentally, the
observed average transverse momentum of a meson increases with the
meson mass.  The average transverse momentum of $\eta$ has however not
been measured.  As the $\eta$ meson has approximately the same mass as
a kaon whose average transverse momentum has been measured, the
average transverse momentum of the $\eta$ meson should be of the order
of the kaon average transverse momentum of $\sqrt{\langle p_T^2
\rangle_K }\sim 0.616$ GeV \cite{Aih88}.  Upon taking this estimate to
be the isoscalar meson average transverse momentum, the experimental
(extrapolated) isoscalar meson ($\eta$ meson) average transverse mass
is
\begin{eqnarray}
\label{scalar}
M_{0T}^h=\sqrt{(0.547)^2+(0.616)^2}~{\rm GeV}\sim 0.824 {\rm ~GeV}.
\end{eqnarray}
We can compare the experimental average transverse masses of isovector
and isoscalar mesons in Eqs.\ (\ref{pionmt}) and (\ref{scalar}) with
the theoretical QCD2 meson masses in two-dimensional space-time in Table
I, which gives $M_{\pi({\rm QCD2)}}$=0.534 GeV, and $M_{\eta({\rm
    QCD2)}}$=0.735 GeV.  We find that there is approximate agreement
of the experimental meson average transverse masses in
four-dimensional space-time with the QCD2 meson masses in
two-dimensional theory, within about 10-15\%.  This approximate
agreement lends support to the identification of the mass of a stable
boson in QCD2$\times$QED2 as the (average) transverse mass of the
boson in four-dimensional space-time.

In the case of QED photons, the rest mass of the photon is zero in
four-dimensional space time.  Hence, the QED2 photon mass in
two-dimensional space-time can be identified as the photon transverse
momentum in four-dimensional space-time.

\section{ Anomalous Soft Photon Transverse Momentum Distribution}

We shall explore how the model of simultaneous meson and photon
production in the string fragmentation process may explain the
anomalous soft photon phenomenon.  In $e^+$-$e^-$ annihilations or
hadron-hadron collisions, $q$-$\bar q$ strings or $q$-(diquark)
strings will be formed with a quark and an antiquark (or a diquark)
pulling apart at the two ends of each string.  As QCD2 mesons and QED2
photons are found to be stable bosons in QCD2$\times$QED2, we can
infer from the quantum field theory of particle production as
described by Casher, Kogut, and Suskind \cite{Cas74} that QCD2 mesons
and QED2 photons will be produced simultaneously in the same process
of $q$-$\bar q$ string fragmentation, when the quark pulls away from
the antiquark (or diquark) at high energies.  The simultaneous
production from the same string explains why anomalous soft photons
are present only in association with hadron production and why the
number of produced mesons and QED2 photons are proportional to each
other, in agreement in the first two features of the anomalous soft
photon phenomenon listed in the Introduction.  We shall examine
whether the QED2 photon model can explain the transverse momentum
distribution in this section and the correlation of the soft photon
yield with hadron production properties in the next section.

\begin{figure} [h]
\includegraphics[angle=0,scale=0.50]{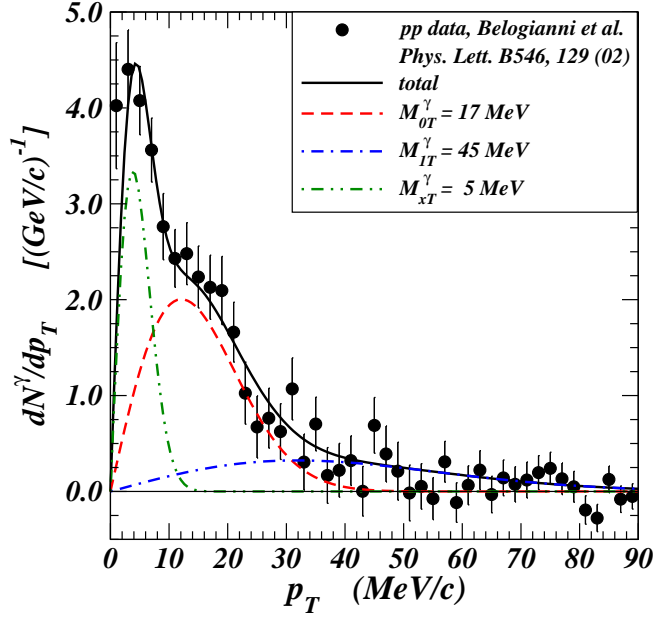}
\vspace*{0.0cm} 
\caption{ (Color Online) Comparison of normalized anomalous soft
  photon $dN^\gamma /dp_T$ per $pp$ collision at 450 GeV/c after
  subtracting the bremsstrahlung contributions \cite{Bel02,Per10} with
  calculated distributions based on three components of anomalous soft
  photons at 17, 45, and 5 MeV.  The solid curve is the total
  theoretical $dN^\gamma /dp_T$ distribution, and the other curves are
  the separate contributions from the three different components.}
\end{figure}

According to the Schwinger's mechanism
\cite{Sch51,Won94a,Wan88,Won95}, the probability of particle
production is an exponential function of the square of the transverse
mass of the produced particle.  For massless photons, the photon
transverse mass is the photon transverse momentum.  It is therefore
reasonable to assume the transverse momentum distribution of each
photon component to be a Gaussian with an average root-mean-squared
transverse momentum $M_{IT}^\gamma$ given by the QED2 photon mass.  In
the measurement of the soft photon transverse momentum distribution in
$e^+$-$e^-$ annihilations, the determination of the orientation of jet
axis has an uncertainty of $\Delta \theta$ $\sim$50 mrad,
corresponding to a root-mean-square uncertainty of $\Delta p_T$
=$E_\gamma \Delta \theta$$ \sim$10 MeV for the small $p_T$ region
\cite{DEL06}.  In the measurement of the transverse momentum
distribution in $pp$ collisions, the uncertainty in angular
measurements is $\Delta \theta$$\sim$10 mrad, which corresponds to
$\Delta p_T$$\sim$2 MeV for the small $p_T$ region \cite{Bel02}.
These uncertainties in the determination of the angles need to be
folded into theoretical calculations in order to compare with
experimental data.  If one assumes a Gaussian distribution of the
transverse coordinates in the angular determination, the folding of
two Gaussian distributions leads to a Gaussian distribution with a
standard deviation square of $[(M_{IT}^\gamma)^2+(\Delta p_T)^2] /2$.
\begin{figure} [h]
\includegraphics[angle=0,scale=0.50]{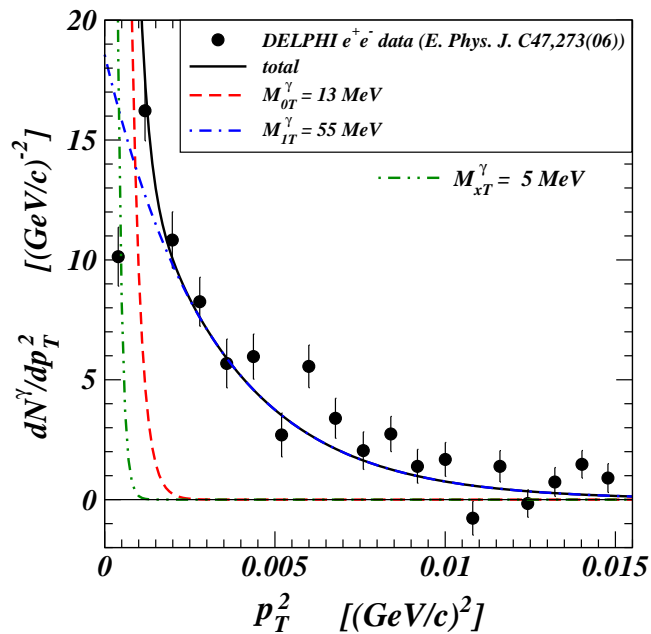}
\vspace*{0.0cm} 
\caption{ (Color Online) Solid points give the DELPHI experimental
  anomalous soft photon $dN^\gamma/dp_T^2$ data in $Z^0$ hadronic
  decay in $e^+$-$e^-$ annihilations, after subtracting the
  bremsstrahlung contributions \cite{DEL06}.  The solid curve is the
  total theoretical $dN^\gamma /dp_T^2$ with contributions from
  component at 13 and 55 MeV, whose separate contributions are shown
  as the dashed and dashed-dot curves, respectively.  The
  dashed-dot-dot curve gives the distribution for the lower momentum
  component at $M_{xT}^\gamma$=5 MeV, which however cannot be resolved
  in the present DELPHI measurements. }
\end{figure}

Based on our theoretical results in Table I, we expect that there will
be two components in the soft photon transverse momentum spectrum.  We
therefore parametrize the experimental anomalous soft photon
transverse momentum distribution as the sum of the two normalized
Gaussian components of isoscalar and isovector photons, each of which
has an average root-mean-squared transverse momentum given by
$(M_{IT}^\gamma)^2+(\Delta p_T)^2$.  We search for distributions
characterized by a transverse mass in the region of
$M_{0T}^\gamma$$\sim$25 MeV for the isoscalar photon and around
$M_{1T}^\gamma$$\sim$44 MeV for the isovector photon component.
However, upon a careful examination of the transverse momentum
distribution of the anomalous soft photons in $pp$ collisions
\cite{Bel02}, we find that in addition to these isoscalar and
isovector photon components, the $pp$ data appears to contain an
additional lower momentum component characterized by a small mass
$M_{{ x}T}^\gamma$.  We need to parametrized it as arising from three
contributions:
\begin{eqnarray}
\frac{dN^\gamma}{dp_T^2} = \sum_{I=0,1}
\frac{N_I^\gamma}{(M_{IT}^\gamma)^2+(\Delta p_t)^2} \exp\left \{ -
\frac{p_T^2}{(M_{IT}^\gamma)^2+(\Delta p_T)^2} \right \}
+
\frac{N_x^\gamma}{(M_{{ x}T}^\gamma)^2+(\Delta p_t)^2} \exp\left \{ -
\frac{p_T^2}{(M_{{ x}T}^\gamma)^2+(\Delta p_T)^2} \right \},
\end{eqnarray}
where the coefficients $N_i^\gamma$ are the integrated numbers of QED2
photons of mass $M_{iT}^\gamma$ produced per event.

The small transverse momentum uncertainties and the extension of the
experimental data down to small values of $p_T$ in $pp$ collisions
make it useful to examine first the transverse momentum distribution
in $pp$ collisions. In Fig. 1 we show the experimental normalized
$dN_\gamma/dp_T$ of anomalous soft photons per $pp$ collision event at
450 GeV/c from Fig.\ 2b of Ref.\ \cite{Bel02,Per10}, after subtracting
the bremsstrahlung contributions.  The $dN^\gamma/dp_T$ data for $pp$
collisions in Fig.\ 1 can be explained by assuming three anomalous
soft photon contributions with parameters
\begin{eqnarray}
\label{fitting1}
M_{0T}^\gamma=17{\rm ~MeV},  
M_{1T}^\gamma=45{\rm ~MeV},  
M_{xT}^\gamma=5 {\rm ~MeV}, 
N_0^\gamma=0.040,  
N_1^\gamma=0.017, 
{\rm ~and~} 
N_x^\gamma  = 0.021. 
\end{eqnarray}
We examine next the anomalous soft photon transverse momentum
distribution in $e^+$-$e^-$ annihilations \cite{DEL06}.  With an error
of $\Delta p_T$ as large as 10 MeV, the distribution cannot be
sensitive to the component at $M_{xT}\sim 5$ MeV.  We show in Fig. 2
the DELPHI experimental $dN^\gamma/dp_T^2$ data from Fig.\ 4f of Ref.\
\cite{DEL06}, after subtracting the inner bremsstrahlung
contributions.  The experimental data can be explained by assuming the
following set of parameters,
\begin{eqnarray}
\label{fitting2}
M_{0T}^\gamma=13{\rm ~MeV},  
M_{1T}^\gamma=55{\rm ~MeV},  
N_0^\gamma  =0.106
{\rm ~and~} 
N_1^\gamma   = 0.058. 
\end{eqnarray}
There are no reliable data points at $p_T^2<0.001$ (GeV/c)$^2$ to fix
the $M_{xT}^\gamma $ component with the present $e^+$-$e^-$ data.  For
illustrative purposes, we show the $M_{xT}^\gamma $ component
calculated with $M_{xT}^\gamma $=5 MeV and
$N_x^\gamma/N_0^\gamma$=0.525 (as in Eq.\ (\ref{fitting1})) shown as
the dashed-dot-dot curve in Fig. 2, to indicate that its presence or
absence has little effects on the theoretical results above
$p_T>0.001$ GeV$^2$.

Our comparison of the anomalous soft photon transverse momentum
distributions reveals that it is necessary to examine the transverse
momentum distributions of both $pp$ collisions and $e^+$-$e^-$
annihilations as complimentary data sets, as the $pp$ data have finer
resolution and smaller errors in the lower $p_T$$\sim$15 MeV regions
while the $e^+$-$e^-$ data have less fluctuations in the higher
$p_T$$\sim$50 MeV region.  The combined analysis indicates that the
transverse momentum spectrum can be qualitative described by two
components with transverse masses of $\sim$15 MeV and $\sim$50 MeV, in
approximate agreement with the gross features of the theoretical QED2
photon model.  There is however an additional, lower momentum
component at $\sim$5 MeV which shows up in $pp$ collisions, but cannot
be resolved in $e^+$-$e^-$ annihilations.  The origin of this
low-$p_T$ component is not known and will be left for future studies.
Among many possibilities, it may be the manifestation of the zero mode
of QED2 photon production whose investigation will be of great future
interest.

\section{Rates of Meson and Anomalous Soft Photon Production}

We shall now examine the remaining feature concerning the rates of
meson and anomalous soft photon production in high-energy $e^+$-$e^-$
annihilations to complete our comparison of the QED2 model with
experimental data.  There are many important physical quantities in
the production processes.  The receding quark and antiquark generate a
QCD field of strength $\kappa_{q\bar q}^h$ and a QED field of strength
$\kappa_{q\bar q}^\gamma$ between the quark and the antiquark in the
flux tube that will produce the QCD mesons and the QED2 photons,
respectively.  Here, we have used the superscript $\alpha$=$h$ for
hadron quantities and $\alpha$=$\gamma$ for photon quantities. Each of
the field quanta is produced in a final state possessing a transverse
momentum, and thus the mass that enters into the consideration of
quanta production should be the transverse mass
$M_{IT}^\alpha=\sqrt{(M_{I({\rm 4D})}^\alpha)^2+p_T^2}$, which can be
identified as the boson mass $M_{I({\rm 2D})}^\alpha$ in
two-dimensional space-time as discussed in Section V.

To obtain an estimate, we can rely on the Schwinger mechanism of
particle production in a strong field as a guide
\cite{Sch51,Won94a,Wan88,Won95}. The probability of particle
production of a composite particle of transverse mass $M_{IT}^\alpha$
depends on the exponential factor of $\exp\{-\pi (M_{IT}^\alpha/2)^2/
\kappa_{q\bar q}^\alpha \}$, where the factor of 1/2 in
$M_{IT}^\alpha/2$ is to denote the production of a pair of particles
each of which has a mass $M_{IT}^\alpha/2$, and the binding of one
particle of mass $M_{IT}^\alpha/2$ with the neighboring particle of
mass $M_{IT}^\alpha/2$ leads subsequently to a composite stable boson
of mass $M_{IT}^\alpha/2+M_{IT}^\alpha/2$.  Furthermore, from
dimensional analysis, we can infer that the rate of production per
space-time volume element $(dz\, dt)$ has the dimension $\kappa_{q
\bar q}^\alpha$.  We therefore assume that the rate of the production
of the number of particle of type $\alpha$, isospin $I$, and mass
$M_{IT}^\alpha$ due to the presence of the QCD and QED fields between
a receding quark $q$ and an antiquark $\bar q$ is
\begin{eqnarray}
\label{rate}
\frac{dN_I^\alpha} {dz\, dt} = A \sum_{q\bar q} P_{q\bar q}
~\kappa_{q\bar q}^\alpha ~\exp \left \{ -\frac{\pi
(M_{IT}^\alpha/2)^2}{\kappa_{q\bar q}^\alpha} \right \},  ~~~\alpha=\gamma,h,
\end{eqnarray}
where $P_{q\bar q}$ is the probability for the
quark-antiquark source pair to be a $u\bar u$ or $d \bar d$ pair, and
$A$ is a dimensionless constant.  In an $e^+$-$e^-$ annihilation at
high energies, there is an equal probability for the quark-antiquark
pair to be a $u \bar u$ or $d \bar d$ pair, and so $P_{u\bar
u}=P_{d\bar d}=1/2$.

For the production of QCD2 mesons, the color electric field strength
between the leading quark and antiquark is independent of the quark
flavor quantum number.  It is given by
\begin{eqnarray}
\label{kappa}
\kappa_{u\bar u}^h=\kappa_{d\bar d}^h={g_{\rm QCD2}^2}/{2}=b.
\end{eqnarray}
For the production of QED2 photons, the electric field strength
between the leading quark $q$ and antiquark $\bar q$ is given in terms
of the electric charges of the quark and the antiquark as
\begin{eqnarray}
\kappa_{q \bar q}^\gamma = |Q_q Q_{\bar q}| e_{\rm QED2}^2/2.
\end{eqnarray}
Thus, we find that
between a receding $q$ and $\bar q$, there is a
constant electric field with a strength
\begin{subequations}
\begin{eqnarray}
\label{uu}
\kappa_{u \bar u}^\gamma = 0.002048 {\rm ~GeV}^2,\\
\label{dd}
\kappa_{d \bar d}^\gamma = 0.000512 {\rm ~GeV}^2.
\end{eqnarray}
\end{subequations}
The QCD field strength $\kappa_{q\bar q}^h$ and the experimental meson
transverse masses as given in Eqs.\ (\ref{pionmt}) and (\ref{scalar})
allow us to determine from (\ref{rate}) the number of mesons (in a
particular $I_3$ state) produced in a space-time volume of $\Delta z
\Delta t$.  Similarly, the QED field strength $\kappa_{q\bar
q}^\gamma$, the QED2 photon isoscalar photon mass of
$M_{0T}^\gamma$$\sim$15 MeV, and the isovector photon mass of
$M_{1T}^\gamma$$\sim$50 MeV from Eqs.\ (\ref{fitting1}) and
(\ref{fitting2}) allow us to determine the number of photons produced.
We obtain,
\begin{subequations}
\label{NN}
\begin{eqnarray}
\label{N1h}
N_1^h &=& A \Delta z \Delta t \times 0.05368 {\rm ~ GeV}^2,\\
\label{N0h}
N_0^h  &=& A \Delta z \Delta t \times 0.01391 {\rm ~GeV}^2,\\
\label{N1g}
N_1^\gamma &=& A \Delta z \Delta t \times 0.0003980 {\rm ~GeV}^2,\\
\label{N0g}
N_0^\gamma &=& A \Delta z \Delta t \times 0.0011206 {\rm ~GeV}^2.
\end{eqnarray}
\end{subequations}
In these estimates, the number of produced particles of different
types and isospin quantum numbers are proportional to the same
space-time volume $\Delta z \Delta t$.  This space-time volume
fluctuates in each $Z^0$ hadronic decay event; the number of mesons
and photons of different isospin quantum numbers will vary from event
to event.  However, because all these particles in each event are
produced simultaneously by the fragmentation of the same $q$-$\bar q$
string and the same space-time volume, the ratio of different isospin
spin quantum numbers and types of particles can therefore be
proportional, on an event-by-event basis. The results in Eq.\
(\ref{NN}) give
\begin{eqnarray}
\frac{N_0^\gamma} {N_1^\gamma} \sim \frac{ 11}{4},   ~~~
\frac{N_0^h}{N_1^h} \sim \frac{1}{4},
\end{eqnarray}
which reveal that isoscalar photons are more preferentially produced
than isovector photons whereas isoscalar mesons are much less likely
produced than isovector meson (in a particular $I_3$ state).  The
theoretical ratio of $N_0^\gamma/N_1^\gamma$=2.8 compares
approximately well with the experimental ratio of
$N_0^\gamma/N_1^\gamma$$ \sim$2.4 and 1.8 in Eq.\ (\ref{fitting1}) and
(\ref{fitting2}) extracted by fitting the experimental transverse
momentum distribution data of \cite{Bel02} and \cite{DEL06} .  Using
these results, we can also construct the ratio of ratios,
\begin{eqnarray}
\frac{N_0^\gamma} {N_0^h} : \frac{N_1^\gamma} {N_1^h} 
\sim \frac{ 11}{4} : \frac{1}{4} = 11 : 1,
\end{eqnarray}
which states that the number of soft isoscalar photons associated with
the isoscalar meson production are more numerous than soft isovector
photons associated with the isovector meson production.

The DELPHI experimental measurements \cite{DEL09,Per09} provide
information on the ratios of anomalous soft photon production with
various produced neutral or charged meson multiplicities.  To compare
with experimental data, we need to convert the number of different
species of mesons to the number of charged and neutral mesons.
Isovector mesons are pions which have two charged states and one
neutral states, and each isoscalar $\eta$ meson decays into 1.64
neutral particles (with 2 $\gamma$'s counted as a $\pi^0$ as in Ref.\
\cite{DEL09}) and 0.57 charged particles.  Thus, the total meson
particle number is $N_{\rm par} = (1.64+0.57) N_0^h+ 3 N_1^h $.  From
Eq.\ (\ref{NN}), the theoretical ratio of total soft photons to total
meson particles (charged and neutral) is
\begin{eqnarray}
\frac{N^\gamma}{N_{\rm par}} 
\sim \frac{N_0^\gamma+N_1^\gamma} {2.21 N_0^h + 3 N_1^h } 
= 7.91 \times 10^{-3}, 
\end{eqnarray}
which compares reasonably well with the experimental ratio of
$N^\gamma/N_{\rm par}\sim 9.1 \times 10^{-3} $.

In our QED2 photon model, production of mesons of isospin quantum
number $I$ will be associated with the production of QED2 photons of
the same isospin quantum number $I$.  Thus, isoscalar QED2 photons
will be associated with isoscalar mesons while isovector QED2 photons
will be associated with isovector mesons.

From Eqs.\ (\ref{N0g}) and (\ref{N0h}), the theoretical ratio of the
number of produced isoscalar photon to the number of produced
isoscalar meson is
\begin{eqnarray}
\label{NN0}
\frac{N_0^\gamma} {N_0^h } = 80.6 \times 10^{-3} .
\end{eqnarray}
As each isoscalar meson produces 1.641 neutral $\pi^0$-like particles
and $0.57$ charged particles, the isoscalar meson is associated with
the production of dominantly neutral mesons.  As we consider the
production of isoscalar mesons to be associated only with the
production of isoscalar photons, then for the isoscalar mode of
production, $N^\gamma/N_{\rm neu} \sim N_0^\gamma/N_{\rm neu}$ which
leads to $N^\gamma/N_{\rm neu} \sim N_0^\gamma / (1.641N_0^h)$ after
summing over all charged particles.  From Eq.\ (\ref{NN0}), the
theoretical ratio of soft photon to neutral particle number is
\begin{eqnarray}
\frac{N^\gamma} {N_{\rm neu} } \sim 
\frac{N_0^\gamma}{1.641N_0^h} =
49.1 \times 10^{-3},
\end{eqnarray}
which comes close to the experimental ratio of $N^\gamma/N_{\rm
neu}\sim 37.7 \times 10^{-3}$.
\begin{table}[h]
  \caption { Comparison of QED2 photon model description of the
  anomalous soft photon production with quantities measured or
  extracted from the $pp$ collision \cite{Bel02} and DELPHI
  $e^+$-$e^-$ annihilation experimental data
  \cite{DEL06,DEL09,Per09}}
\vspace*{0.2cm} 
\hspace*{-0.0cm}
\begin{tabular}{|c|c|c|}
\hline Quantities & QED2 Model & Experimental Anomalous Soft Photon
Data \\ & & \\ \hline 
Isoscalar photon mass  $M_0$ & $O( 25$ MeV) & $\sim$15 MeV \\ \hline 
Isosvector photon mass $M_1$ & $O( 44$ MeV) & $\sim$50 MeV \\\hline 
$N_0^\gamma/N_1^\gamma$ & 11/4 & 1.8-2.4 \\ \hline
$N^\gamma/N_{\rm par}$ & 7.91$\times 10^{-3}$ &  9.1$\times 10^{-3}$ \\\hline 
$N^\gamma/N_{\rm neu}$ & 49.1$\times 10^{-3}$ & 37.7$\times 10^{-3}$ \\ \hline 
$N^\gamma/N_{\rm ch}$  & 3.71$\times 10^{-3}$ &  6.9$\times 10^{-3}$ \\ \hline
\end{tabular}
\end{table}

From Eqs.\ (\ref{N1g}) and (\ref{N1h}), the theoretical ratio of the
number of produced isovector photon to the number of produced
isoscalar meson in a particular $I_3$ state is 
\begin{eqnarray}
\label{NN1}
\frac{N_{1}^\gamma} {N_{1}^h } = 7.41 \times 10^{-3} .
\end{eqnarray}
The isovector meson is three-fold degenerate with two charged
particles and one neutral particle. Thus, the production of an
isovector meson is associated with the production of dominantly
charged particles.  In the QED2 photon model, the sources of isospin
current disturbances that produce the mesons and photons are the
same.  Therefore, the production of isovector mesons is associated
only with the production of isovector photons. Consequently, we have
$N^\gamma/N_{\rm ch} \sim N_1^\gamma/N_{\rm ch}$, which leads to
$N^\gamma/N_{\rm ch} \sim N_1^\gamma / (2N_1^h)$ after summing over
all neutral particles.  Equation (\ref{NN1}) then
leads to
\begin{eqnarray}
\frac{N^\gamma} {N_{\rm ch} } \sim
\frac{N_1^\gamma} {(2 N_1^h)}
= 3.71 \times 10^{-3} , 
\end{eqnarray}
which is slightly less than the experimental ratio $N_\gamma/N_{\rm
  ch} \sim 6.9 \times 10^{-3}$, but is within the same order of
magnitude.

As a summary, we give the comparisons of various quantities obtained
in the QED2 model with the DELPHI data
\cite{DEL06,DEL09,Per09} in Table II.  We conclude from the
comparison that gross features of the DELPHI data is approximately
consistent with the QED2 photon model.

\section{Further Experimental Tests of the QED2 Photon Model}

While the QED2 photon model appears to explain qualitatively the main
features of the experimental anomalous soft photon data, it is
desirable to carry out further experimental measurements to test the
model:

\begin{enumerate}

\item It will be of interest to measure the transverse momentum
  distribution of the soft photons with a finer $p_T$ resolution and
  greater precision for a given narrow range of photon rapidities.
  Qualitatively, we expect the production of photons with two
  different average transverse momenta, one at $\sim$15 MeV for the
  production of the isoscalar photon and one at $\sim$50 MeV for the
  production of the isovector photon.

\item It will be necessary to confirm the presence of the low momentum
  component at $M_{xT}^\gamma$=5 MeV in high-energy $e^+$-$e^-$
  annihilation experiments. As the origin and the properties of such a
  low $p_T$ component is still unknown, additional experimental
  information on this source of anomalous soft photons will improve
  our understanding of such a component.

\item It will be of interest to measure the transverse momentum
  distribution by selecting events with predominantly neutral
  particles and events with predominantly charged particles.  The
  former events will likely arise from the production of isoscalar
  mesons and QED2 isoscalar photons, with an average photon transverse
  momentum of $\sim$15 MeV, while the latter from the production of
  isovector mesons and the isovector photons, with an average photon
  transverse momentum of $\sim$50 MeV.

\item
The rapidity distribution of the produced photons should exhibit the
plateau structure, as expected of similar distributions in meson
production.  A measurement of the rapidity distribution will provide
useful additional information on the dynamics of soft photon production.

\item Measurements of the properties of associated hadrons similar to
  those of the DELPHI Collaboration should be carried out with
  hadron-hadron collisions at high energies where anomalous soft
  photon production has been reported
  \cite{Chl84,Bot91,Ban93,Bel97,Bel02,Bel02a}.

\end{enumerate}

\section{Conclusions and Discussions}

A color flux tube is formed when a quark and an antiquark (or a
diquark) pull apart from each other at high energies.  The motion of
the quarks in the underlying vacuum of the flux tube generates color
charge oscillations which lead to the production of mesons.  As a
quark carries both a color charge and an electric charge, the color
charge oscillations of the quarks in the vacuum are accompanied by
electric charge oscillations, which will in turn lead to the
simultaneous production of soft photons during the meson production
process.

To study these density oscillations, we start with quarks interacting
with both QCD and QED interactions in four-dimensional space-time in
the U(3) group which breaks into the color SU(3) and the QED U(1)
subgroups.  Specializing to particle production at high energies, we
find that the dominance of the longitudinal motion and transverse
confinement lead to the compactification from QED4$\times$QED4 in
four-dimensional space-time to QCD2$\times$QED2 in two-dimensional
space-time, with the formation of the flux tube.  In the flux tube, the
self-consistent coupling of quarks and gauge fields lead to color
charge and electric charge oscillations that give rise to stable QCD2
bosons and QED2 bosons.  The boson masses depend on the gauge field
coupling constants.  The presence of the flavor degrees of freedom
leads to isospin dependence of the boson masses, with the isovector
meson mass smaller than the isoscalar meson mass, but the mass
ordering is reversed for the isoscalar photon and the isovector
photon.

As QCD2 and QED2 bosons are stable in the flux tube environment, we
can infer from the quantum field theory description of particle
production in Ref.\ \cite{Cas74} that these QCD2 mesons and QED2
photons will be produced simultaneously in $q$-$\bar q$ string
fragmentation.  Under the condition of adiabaticity with no change of
the particle energy and longitudinal momentum after the produced
particle emerges from the production region, the boson mass in
two-dimensional space-time turns into the boson transverse mass in
four-dimensional space-time.

The QED2 photon model can explain various features of the anomalous
soft photon phenomenon.  Because both color charge oscillations and
electric charge oscillations arise from the same density oscillations
of the quarks in the vacuum, both QCD2 meson and QED2 photon will be
simultaneously produced by the fragmentation of the $q$-$\bar q$
string.  These features are in agreement with those observed in DELPHI
experiments \cite{DEL08,DEL09,Per09}.  The transverse momentum
distributions of anomalous soft photons in $pp$ collisions
\cite{Bel02} and $e^+$-$e^-$ annihilations \cite{DEL06} can be
described by a component with $M_{0T}^\gamma$$\sim$15 MeV and a
component at $M_{1T}^\gamma$$\sim$50 MeV in approximate agreement with
theoretical estimates of the order of the isoscalar and isovector QED2
photon masses.

In the QED2 model, there are important and non-trivial isospin
dependencies in the rate of photon and hadron productions that is
consistent with recent DELPHI data.  The model predicts that the
isoscalar photon mass is lower than the isovector photon mass.
Consequently, the production of isoscalar photons is more likely than
isovector photons.  In contrast, the QCD isoscalar meson mass is
greater than the isovector mass, the production of isoscalar mesons is
less likely than isovector mesons.  Thus, the ratio of
$N_0^\gamma/N_0^h$ can be much greater than $N_1^\gamma/N_1^h$.  The
production of isoscalar hadrons is associated with the production of
isoscalar photons and leads predominately to neutral particles while
the production of isovector hadrons is associated with the production
of isovector photons and leads predominantly to charged particles.  As
a consequence, the ratio $N^\gamma/N_{\rm neu}$ is much greater than
the ratio $N^\gamma/N_{\rm ch}$, as observed by the DELPHI
Collaboration \cite{DEL09,Per09}.

Although the QED2 photon model appears to be promising, it is
desirable to carry out additional experimental measurements to test
the model.  We suggest the search for the two components of transverse
momentum distributions by making appropriate cuts in soft photon
rapidities and selecting different regions of neutral and charge
multiplicities where different isospin photon components are expected.
The identification of the two components of different soft photon
transverse momenta will be a crucial test of the QED2 photon model in
the QCD string fragmentation process.

Our examination of the transverse momentum distribution of anomalous
soft photons in $pp$ collisions in \cite{Bel02} reveal the presence of
an additional component characterized by a transverse mass of
$M_{xT}^\gamma$=5 MeV.  What is the nature of this component?  Is it
related to the zero mode of density oscillations?  How does the zero
mode manifest itself experimentally?  Experimental investigation of
the low transverse momentum component of photon production and the
theoretical investigation of the zero mode of QED2 photon production
will be of great future interest.

There are puzzling elements of the QED2 photon model that call for
future theoretical and experimental resolution.  As it now stands, the
theoretical determination of the QED2 photon masses is rather
uncertain as the mass scale $\mu$ in the bosonization of the scalar
density is unknown.  The QED2 photon mass scale as extracted from
experimental data requires an electromagnetic current quark mass
smaller than the current quark mass as determined from perturbative
QCD.  Whether or not such a smaller value of the mass scale $\mu$ for
the anomalous soft photon production is justified will require further
theoretical and experimental investigations.

Another puzzling and unresolved question is the more detail
description of the evolution from a QED2 photon to a QED4 photon.  We
have used the concept of adiabaticity so that the photon preserves its
energy and longitudinal momentum, only to develop a transverse
momentum to balance the mass shell condition.  Such a description
appears to give a qualitative description of the transverse momenta
and production probabilities of the soft photons. However, there is no
additional content in the dynamics of the evolution in our hypothesis.
A more detail dynamics of the evolution of QED2 to QED4 will be of
great interest.

Finally, if the model is proved to be successful in explaining the
anomalous soft photon data, it may be useful to explore whether one
can study this non-perturbative problem in the full four-dimensional
space-time without resorting to the intermediate stage of going through the
two-dimensional space-time, where non-perturbative physics can be
carried out more readily.  The success of a completely
four-dimensional description will provide new insight into the
non-perturbative behavior of particle production in strong fields.

\null
\vskip 0.5cm
\centerline{\bf Acknowledgment}
\vskip .5cm
The author would like to thank Dr.\ V.\ Perepelitsa for stimulating
discussions and valuable information on DELPHI anomalous soft photon
experimental data.  The author also wishes to thank Drs.\ H.\ Crater
and T.\ Barnes for helpful discussions.  The research was sponsored by
the Office of Nuclear Physics, U.S. Department of Energy.


\begin{thebibliography}{200}
\bibitem{noteH} Compared to the result of Ref.\ \cite{Col76}, Eq.\
(\ref{hh}) contains an extra factor of 1/2 which arises in the scale
we have chosen to define $\phi^\alpha$ with Eq.\ (\ref{uua}), where
tr$(t^\alpha t^\beta)$=$\delta_{\alpha \beta}/2$ for non-Abelian
bosonization.  If we choose a differently scaled boson field with
$u=\exp\{ i \sqrt{8\pi} (\phi')^\alpha t^\alpha \}$, then there will
be no extra factor of 1/2 in the Hamiltonian density with the
$(\phi')^\alpha$ fields.
\bibitem{notemass} The mass of the massive boson in multiflavor QCD2
  was given as $M$=$g_{QCD2}\sqrt{N_f/2\pi}$ in \cite{Arm99,Abr04} but
  as $M$=$g_{QCD2}\sqrt{N_f/\pi}$ in \cite{Tri02,Abr05}.  In our
  comparison, we shall quote only the latest result,
  $M$=$g_{QCD2}\sqrt{N_f/\pi}$ of Ref.\ \cite{Abr05}, which comes from
  the same group of authors and presumably has included the latest
  typographical and other corrections.
\end{thebibliography}
\end{document}